\documentclass[twocolumn,epsf,prb,amsmath,amssymb,showpacs]{revtex4}
\usepackage{epsfig}
\usepackage{graphicx}
\usepackage{epstopdf}
\usepackage{amsmath}
\usepackage{amssymb}
\usepackage{color}

\DeclareMathAlphabet{\bi}{OML}{cmm}{b}{it}

\begin{document}
\title{Spectroscopy of snake states using a graphene Hall bar}

\author{S. P. Milovanovi\'{c}}\email{slavisa.milovanovic@gmail.com}
\affiliation{Departement Fysica, Universiteit Antwerpen, \\
Groenenborgerlaan 171, B-2020 Antwerpen, Belgium}

\author{M.~Ramezani Masir}\email{mrmphys@gmail.com}
\affiliation{Departement Fysica, Universiteit Antwerpen, \\
Groenenborgerlaan 171, B-2020 Antwerpen, Belgium}

\author{F.~M.~Peeters}\email{francois.peeters@ua.ac.be}
\affiliation{Departement Fysica, Universiteit Antwerpen, \\
Groenenborgerlaan 171, B-2020 Antwerpen, Belgium}

\begin{abstract}
An approach to observe snake states in a graphene Hall bar containing a pn-junction is proposed. The magnetic field dependence of the bend resistance
in a ballistic graphene Hall bar structure containing a tilted
pn-junction oscillates as a function of applied magnetic field. We
show that each oscillation is due to a specific snake state that
moves along the pn-interface. Furthermore depending on the value of
the magnetic field and applied potential we can control the lead in
which the electrons will end up and hence control the response of
the system.
\end{abstract}
\maketitle

Graphene's electronic properties are drastically different from
conventional semiconductors. Graphene has a linear spectrum near the
$K$ and $K'$ points with zero gap\cite{f1,f2} which causes perfect
transmission through arbitrarily high and wide barriers for normal
incidence, referred to as Klein tunneling \cite{f3,f4,f5,f6,f7}. The
metamaterial character of pn-junctions in graphene\cite{chei07} was
pointed out earlier, and focusing of electronic waves was
proposed\cite{Matu1,has10}. The metamaterial properties of the above
mentioned pn-junctions resulted in the expectancy of controlling the
electron wave function, in particular, the width of electron beams
by means of a superlattice  that is known as collimation
\cite{Cheol}. Qualitatively, the metamaterial properties of
pn-junctions in graphene can be understood by inspecting classical
trajectories\cite{A2}, or using ray optics as it is called in the
case of electromagnetic phenomena\cite{A1}. Classical simulations
for electronic transport were done recently for a Hall bar made of
single layer\cite{A3, cMLG} and bilayer\cite{AM2} graphene. Gapless
energy dispersion of graphene allows electron and hole switching at
the pn-interface which can be realized using nanostructured gates
\cite{mrm1,mrm2}.

Applying a nanostructured top gate or side gates
one can induce a pn-interface in the Hall cross as shown
schematically in Fig. 1(a). Near the interface conduction by
electrons on one side and conduction by holes on the other side
occurs. An applied magnetic field bends electron and hole
trajectories towards the interface while Klein tunneling through the
interface allows snake orbits to propagate (see Fig. 1(b)). Snake
states along the pn-interface were predicted analytically
\cite{Ma1,Ma2} in the presence of a homogeneous magnetic field.
Existence of such states was confirmed in recent experiments
\cite{Marcus1,Marcus0} by measuring the resistance along the
pn-interface.

Here we showed that by using a tilted pn-interface in the cross of a Hall
bar structure allows us to characterize the
different snake states by measuring the bend resistance. In such a
device carriers injected from any lead ( Fig. 1(c) shows injection
from lead 1) will transmit multiple times on the pn-interface and
move along it until they reach one of the leads at the other end
(lead 2 or lead 3 in Fig. 1(c)). The choice of the final lead
depends on the value of applied magnetic field, carrier density,
length of the pn-interface and the angle of injection. We found that as a
function of the magnetic field strength or the Fermi energy a
sequence of peaks and dips appear in the bend resistance depending
in which lead the carriers will end up. This effect can also be
viewed as a type of magnetic focusing, but unlike normal transverse magnetic
focusing\cite{TMF} which is a result of skipping orbits, here the
focusing appears due to snake states.
\begin{figure}[h!]
  \centering
  \includegraphics[width=7cm]{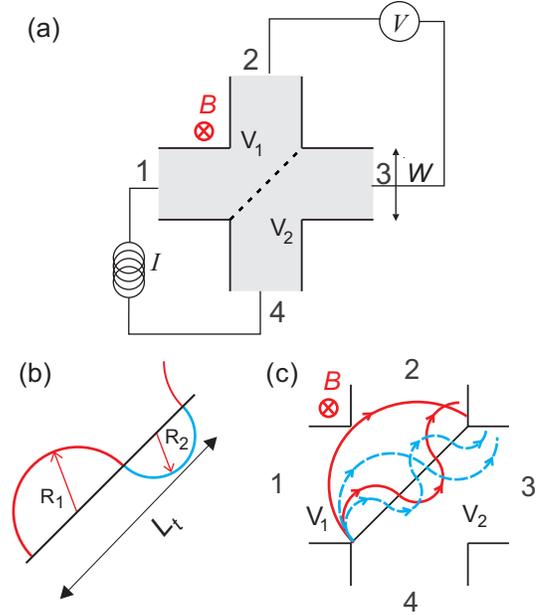}
  \caption{(a) Schematics of the Hall bar with
  tilted pn-junction. (b) Snake states through the pn-interface in the presence of
  the perpendicular magnetic field. (c) Four possible trajectories for an electron
  injected from lead 1.}
  \label{fig1}
\end{figure}
\begin{figure}[h!]
  \centering
  \includegraphics[width=8cm]{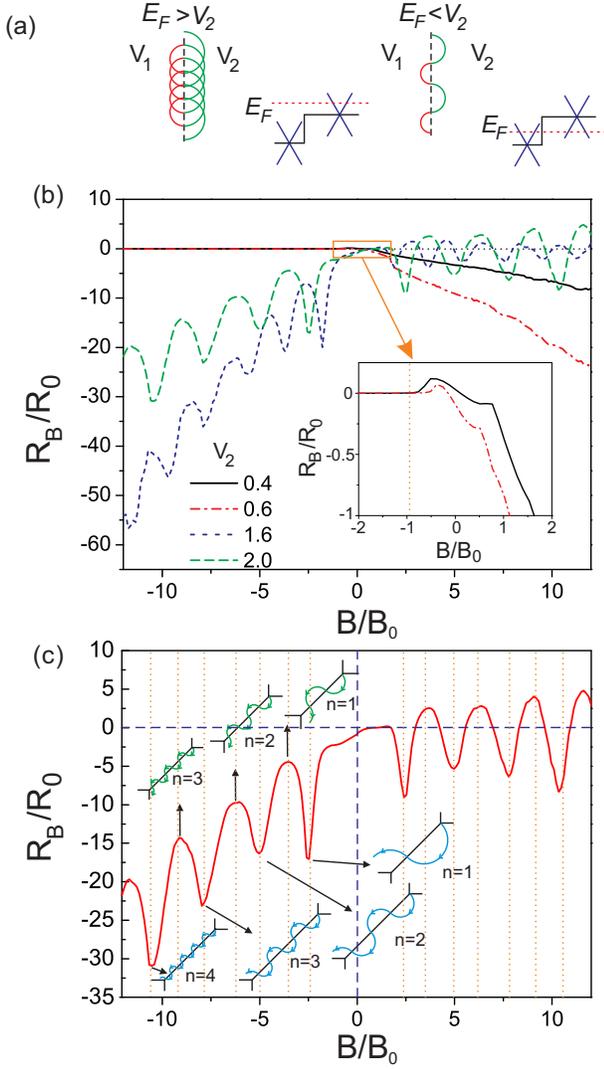}
  \caption{(Color online)(a) The schematics of two possible
  trajectories
  for two different values of the Fermi energy. (b) The bend resistance for different
  values of applied potential $V_{2}$ with $V_{1} = 0$ and $E_{F}/E_{0} = 1$.
The inset shows a zoom around low magnetic field.
  (c) Bend resistance for the case $V_{2}/E_{0} = 2$ and $E_{F}/E_{0} =
1$. The different snake trajectories for some peaks and dips are
shown in the insets.}
  \label{fig2}
\end{figure}
We can control the position of the peaks by changing the cyclotron
radius which is given by,
\begin{equation}\label{RC}
    R_{c} = \displaystyle{\frac{|E_{F} - V|}{e v_{F}|B|} = \frac{\hbar \sqrt{\pi n_s}}{e |B|}
    }
\end{equation}
where $n_s$ is the carrier density, $E_{F}$ is the Fermi energy, $V$
is the applied potential, $v_{F}$ is the Fermi velocity and $B$ is
the applied magnetic field. Thus we are able to control the snake states in
graphene by changing the magnetic field or carrier density.

To simulate the transport properties of such a graphene Hall bar we
rely on the semiclassical billiard model\cite{cBM}. In this model
electrons are considered as point particles (billiards) which are
injected uniformly over the length of the lead, while the angular
distribution is given by $\displaystyle{P(\alpha) =
\frac{1}{2}\cos(\alpha)}$, with $\alpha \in [-\pi/2,\pi/2]$. The
model is justified when $\lambda_F\ll W$, where $\lambda_F$ is the
Fermi wavelength and $W$ the width of the lead and when quantization
effects are not important. This approach has been used
to describe various experiments with a mesoscopic Hall
bar\cite{A3,cBM,cSCA,cSCA4,cSCA5}. The motion of ballistic particles is
determined by the classical Newton equation of motion, which is
justified for the case $l_\phi<W<l_e$ where $l_\phi$ is the phase
coherence length and $l_e$ the mean free path (for typical
parameters at low temperatures the electron mean free path can be
calculated as $l_e=(\hbar /e)\mu (\pi n_s)^{1/2}> 1\mu m$, with
$\mu$ the mobility and $n_s$ the electron density), while the
transmission of electrons and holes through the pn-interface is
calculated quantum mechanically using the Dirac Hamiltonian.

Transport properties of the system are obtained by using the
Landauer-B\"{u}ttiker formalism. For this purpose we need to find
the electron transmission probabilities between the different leads
of the Hall bar structure. The probability that an electron injected
from terminal $j$ will end up in terminal $i$ is given by $T_{ij}$.
These transmission probabilities are then used in the
Landauer-B\"{u}ttiker formula in order to calculate the current in
terminal $i$,
\begin{equation}
 I_i = \frac{e}{h}\left[(N_i-T_{ii})\mu_i-\sum_{j\neq i}T_{ij}\mu_j\right],
\end{equation}
\begin{figure}[h!]
  \centering
  \hspace*{-0.5 cm}
  \includegraphics[width=9.0cm]{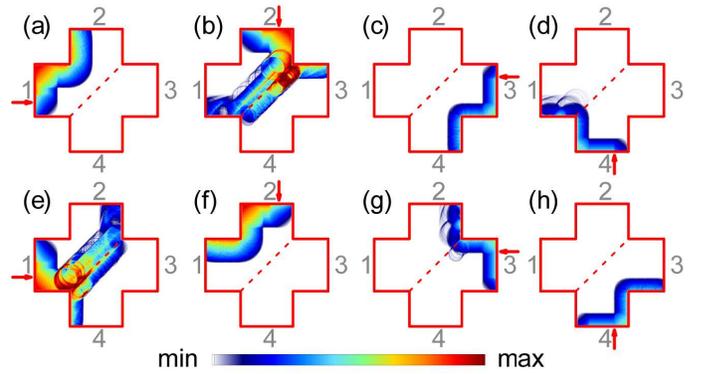}
  \caption{(Color online) Electron current density for $V_{1} =
  0$, $V_{2}/E_{0}=0.4$ and $E_{F}/E_{0} = 1$, when (a) - (d) $ B/B_{0} = -3$ and
  (e) - (f) $B/B_{0} = 3$. The arrow indicates the lead at which the current is injected.}
  \label{fig2}
\end{figure}
here $N_i$ is the number of occupied transport channels, which
depends on $E_F$ and $\mu_i$ and $\mu_j$ are the chemical potentials of
the reservoirs $i$ and $j$, respectively,  $e$ is the electron
charge and $h$ is Planck's constant. Eliminating the chemical
potentials we can derive expressions for the different resistances,
\begin{equation}\label{RB0}
 R_{mn,kl} = \frac{h}{e^2}\frac{T_{km}T_{ln}-T_{kn}T_{lm}}{D},
\end{equation}
with $D =\alpha_{11}\alpha_{22}-\alpha_{12}\alpha_{21}$ and
\begin{equation}
\begin{array}{l}
\alpha_{11} = \left[(N_i-T_{11})S-(T_{14}+T_{12})(T_{41}+T_{21})\right]/S \\
 \alpha_{12} = (T_{12}T_{34}-T_{14}T_{32})/S \\
   \alpha_{21} =(T_{21}T_{43}-T_{41}T_{23})/S \\
 \alpha_{22} = \left[(N_i-T_{22})S -(T_{21}+T_{23})(T_{32}+T_{21})\right]/S,
\end{array}
\end{equation}
where $ S = T_{12}+T_{14}+T_{32}+T_{34}$. In the present paper, we
are interested in the bend resistance $R_B = R_{14,32}$.\\

The four-terminal Hall bar with a tilted pn-interface is used as the device of interest. For
numerical calculations we used $E_{F} = 50$ meV, $V_{1} = 0$ and
$V_{2} = 100$ meV. For a typical electron density $n_s = 1.84
\times 10^{11} cm^{-2}$, and a width of leads $W = 1 \mu m$, $100
\mu m$ and $10 nm$, we obtain for the unit of magnetic field $B_{0}
= |E_{F}|/ev_{F}W= 0.05T$, $0.5$ and $5 T$ and for the resistance unit $R_0
= (h/4e^2)(\hbar v_F/E_FW) = 0.085k\Omega$, $8.5 k\Omega$ and
$0.85k\Omega$. The cyclotron radius is given by Eq. (\ref{RC})
which for $B_{0}$ results into $R_{c0} = W = |E_{F}|/ev_{F}B_{0}$.
The numerical simulation of the bend resistance as a function of the
magnetic field is shown in Figs. 2(b) and 2(c). Two different
regimes are found:

1) $ E_{F} > V_{2}$: the electron can pass through the pn-interface
and preserves the direction of motion with a change of the cyclotron
radius. One of the possible trajectories is shown in the left panel
of Fig. 2(a). Notice that from Eq. (\ref{RB0}) the bend resistance
is proportional to
\begin{equation}\label{RB}
    R_{B} \propto \underbrace{T_{31}T_{24}}_{I} -\underbrace{T_{34}T_{21}}_{II}.
\end{equation}
As shown in Fig. 2(b) for negative magnetic field the bend resistance
is almost zero. We can understand this behavior better using the
electron current density plots, shown in Figs. 3(a)-(d) for
$B/B_0=-3$. The electron current density for electrons injected from
lead 1 (see Fig. 3(a)) shows that the majority of electrons perform
skipping orbits on the edge of the system and therefore end up in
lead 2 resulting in zero transmission probability $T_{31}$ in part
$I$ of Eq. (\ref{RB}). Similarly, shown in Fig. 3(d), none of the
electrons travel from lead 4 to lead 3, resulting in zero $T_{34}$
in part $II$ of Eq. (\ref{RB}). Consequently, for high negative magnetic
fields the bend resistance is zero. As we approach zero
magnetic field the cyclotron radius increases and the electrons will
have a chance to travel from lead 1 to 3 or from lead 4 to lead 3.
We can find this classically by setting the cyclotron radius to
$R_{c} = D/2 = \sqrt{2}W/2$ with corresponding magnetic field given
by $B/B_{0} = \sqrt{2}(1-V/E_{F})$ (see the inset of Fig. 2(c) for
the position of the classically predicted value of the magnetic
field of $B/B_{0} = 0.84$ given by the vertical orange dotted line
for $V_{2}/E_{F} = 0.4 $). For positive magnetic field electrons
coming from lead $1$ travel through the pn-junction with different
radius on the two sides of the junction as shown schematically in
the left panel of Fig. 2(a) which has a chance to end up in any lead
and consequently we have nonzero transmission factors $T_{i1}$
(especially $T_{31}$ and $T_{21}$ which appear in Eq. (\ref{RB})).
On the other hand, all electrons injected from lead $4$ perform
skipping orbits and consequently will end up in lead $3$ resulting
in a nonzero $T_{34}$. However the transmission $T_{24}$ appearing
in Eq. (\ref{RB}) is zero because there is no electrons going from
lead $4$ into $2$, then the first part of Eq. (\ref{RB}) is zero and
only the second part will be nonzero and $R_{B} \propto -
T_{34}T_{21}$ which is responsible for the negative bend resistance.

2) For $E_{F} < V_{2}$ we have on one side of the pn-junction
electrons and on the other side holes. When an electron transmits
through the pn-interface it transforms to a hole and its direction
of motion will change which results in a snake state, as shown in
Figs. 1(b)-(c) and in Fig. 2(a). Depending on initial factors (angle
of injection, magnetic filed, potential, etc...) it can end up in the
electron or hole region. This results in an oscillatory behavior of
the bend resistance as shown in Figs. 2(b) and 2(c). Each of the
peaks can be associated with a specific snake state as illustrated
schematically in the inset of Fig. 2(c). Peaks occur respectively
when the length of the tilted pn-interface ($D = \sqrt{2} W$)
satisfies the equality,
%
%$\displaystyle{D^{dips} = {\left(2n - 1\right)R_{1} + 2nR_{2}}}$ and
$\displaystyle{D^{peaks} = {2nR_{1} + \left(2n +1\right)R_{2}}}$
which means that the particle injected from lead $m(n)$ will most
likely end up in lead $l(k)$. Here $n$ is the number of times the
particle switches between the electron and the hole region along the pn-interface.
The corresponding magnetic field is given by,
\begin{equation}\label{E3}
\begin{array}{c}
  %\displaystyle{B^{peaks}_{n}  = \frac{B_{0}}{\sqrt{2}} [2n(V/E_{F}) -1], ~~ n = 1,2,...}, \\
  \displaystyle{B^{peaks}_{n}  = \frac{B_{0}}{\sqrt{2}} [2n(V/E_{F}) +1] , ~~ n =1,2,...}.
\end{array}
\end{equation}
where we use $V_{1} = 0$ and $V_{2} = V$. A contour plot of the bend
resistance as a function of applied magnetic field and electrical
potential is plotted in Fig. 4. Figure shows reasonable good
agreement between classically predicted peaks (Eq. ({\ref{E3})),
shown by the black dashed lines, and our simulations. For the special
case when $R_{1} = R_{2}$ (or $V_{2}/E_{F} = 2$) we find
%
%%
%\begin{equation}\label{E1}
%    \begin{array}{c}
%      L^{dips}_{t} = (4n - 1)R_{c},~n = 1,2, ... \\
%      L^{peaks}_{t} = (4n + 1)R_{c},~n = 1,2, ...
%    \end{array}
%\end{equation}
%%
%
\begin{figure}[h!]
  \centering
  \hspace*{-1 cm}
  \includegraphics[width=7cm]{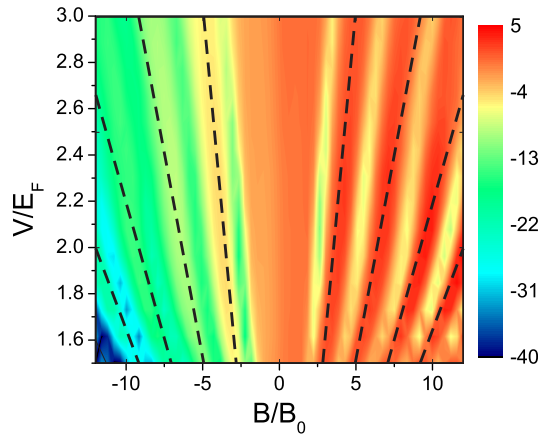}
  \caption{(Color online) Bend resistance as a function of magnetic field and
  applied potential. The black dashed lines correspond to classically predicted peaks
  given by Eq. (\ref{RB}).}
  \label{fig2}
\end{figure}
\begin{figure}[h!]
  \centering
  \hspace*{-0.5cm}
  \includegraphics[width=9.0cm]{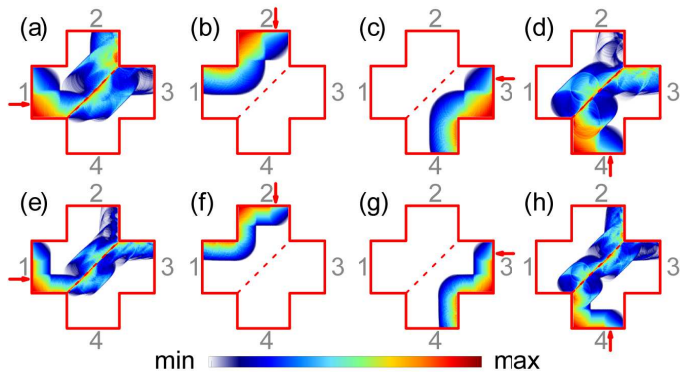}
  \caption{(Color online) Electron densities for $V_{1} =
  0$, $V_{2}/E_{0}=2$ and $E_{F}/E_{0} = 1$, (a) - (d) when $B/B_{0} = 2.4$ and
   (e) - (f) when $B/B_{0} = 3.5$. The arrow indicates the lead at which
the current is injected.}
  \label{fig2}
\end{figure}
that the position of the peaks is given by, $B^{peaks}_{n}  =
B_{0}(4n+1)/\sqrt{2}$. The analytical results of Eq. (\ref{E3}) are
shown in Fig. 2(c) by the vertical dotted orange lines. We also show
schematically the corresponding snake trajectory for each peak and
dip for different number of nodes.
%
%%
%\begin{equation}\label{E4}
%    \displaystyle{B^{dips}_{n}  = \frac{B_{0}}{\sqrt{2}}(4n-1) , ~~~~~~~ n =
%    1,2,...}, %= \left(\frac{|E_{F} - V_{1}|}{eL_{t}v_{F}}\right)(4n-1)
%\end{equation}
%%
%For these parameters the cyclotron radius on both sides of the
%tilted pn-interface are the same. We show these analytical values
%with straight line in Fig. 2(b) and they are in good agreement with
%numerical result.
%The positions of the dips and
%peaks are given in the following table:\\
%\\
%%
%\begin{tabular}{l*{4}{c}}
%n              & 1 & 2 & 3 & 4     \\
%\hline
%$B^{dips}/B_{0}$    & 2.12 & 4.95 & 7.78 & 10.6   \\
%\hline
%$B^{peaks}/B_{0}$    & 3.54 & 6.36 & 9.2 & 12  \\
%\end{tabular}
%%
%\\
The distance between consecutive peaks is given by
\begin{equation}\label{Del}
    \displaystyle{\Delta B = B_{n+1} - B_{n} = 4 \left(\frac{|E_{F} -
    V_{1}|}{e L_{t} v_{F}}\right),}
\end{equation}
which for the used parameters results into $\Delta B = 2\sqrt{2}
B_{0} \approx 2.83 B_{0}$. In Figs. 5(a)-(d) we plot the electron
current densities for $B/B_{0} = 2.4$ which corresponds to a dip in
the bend resistance.  Figures show that the majority of carriers
injected from lead 1 will end up in lead 2. In case of injection
from lead 4, carriers are most likely to end up in lead 3. Other
transmission coefficients $T_{31}$ and $T_{24}$ are small which
result in a minimum in the bend resistance. For the magnetic field
$B/B_{0} = 3.54$, which corresponds with a peak,
most electrons injected from lead $1$ will end up into lead $3$ and
electrons injected from lead $4$ will end up into lead $2$ which
means that part I in Eq. (\ref{RB}) will be much larger than part
II.

In summary, using a tilted pn-interface we are able to selectively
probe different snake states and investigate its influence on the
transport properties of a graphene Hall bar. Such a pn-junction along the diagonal of the Hall bar cross can be
realized experimentally. As our numerical simulation showed, applying
different magnetic field (or different potential) we are able to
control the electron current along the pn-interface. This resulted
in an oscillatory behavior of the bend resistance which is a
signature of the different snake states appearing in the system. We
found an analytical formula that predicts the position of the peaks
and dips in the resistance.

This work was supported by the Flemish Science Foundation (FWO-Vl),
the European Science Foundation (ESF) under the EUROCORES Program
EuroGRAPHENE within the project CONGRAN and the Methusalem
Foundation of the Flemish government.

\end{document}